\documentclass{cernrep} 
\usepackage{texnames}
\usepackage[T1]{fontenc}
\begin{document}
\title{Basics of Feature Selection and Statistical Learning for High Energy Physics}
 
\author{Anselm Vossen}

\institute{Albert-Ludwigs-Universit\"at, Freiburg, Germany}

\maketitle 

\begin{abstract}
This document introduces basics in data preparation, feature selection 
and learning basics for high energy physics tasks. The emphasis is on feature selection by principal
component analysis, information gain and significance measures for features.
As examples for basic statistical learning algorithms, the maximum a posteriori and maximum likelihood classifiers
are shown. Furthermore, a simple rule based classification as a means for automated cut finding is introduced.
Finally two toolboxes for the application of statistical learning techniques are introduced.
\end{abstract}
 
\section{Introduction}
Modern methods in data analysis for High Energy Physics (HEP) usually rely on learning algorithms 
such as neural networks. 
The popularity of the latter has led to many out of the box solutions that are easy to use and widely employed
in the physics community (e.g. the ROOT based package TMVA \cite{tmva}). 
However, it is often ignored that even the most advanced algorithms strongly depend on the quality of the input.
If the input is poorly prepared, the performance will suffer as well. 
This lecture will present some basic means to prepare the input and evaluate the value of a feature.
When the input is well prepared good classification results with simple classification schemes can often be achieved,
some of which will be introduced here such as Bayesian schemes and rule based classification. 
The advantages of using a simple algorithm is, that the resulting functions are easy to interpret,
 compared with complex schemes like neural networks or support vector machines. 
For example results of a rule based classification can be used and interpreted as a normal cut based analysis.
This lecture will also present overtraining
and performance evaluation as points that are common to all classification tasks. 
All of these are done with a special regard to the specifics of problems that are encountered in particle physics:
outliers in the data caused by noise, training data not readily available and produced by simulations 
which are not always accurate. 
For further reading \cite{patClas} and \cite{kltCite} are recommended.

\section{Example, nomenclature}

A typical problem at the track level in particle physics where statistical learning schemes are nowadays routinely employed
consists of the classification of particles.
The input to a classification algorithm is made up of an ordered set of attributes
which characterize the event that should be classified. The terms feature and attribute may 
be used interchangeably. 
The features in particle classification problems usually consist of measurable quantities 
like energy loss and particle momentum. 
In addition features that incorporate prior knowledge about the process
are important. 
This information usually follows from constraints given by the laws of physics.

\subsection{Incorporating prior knowledge}

Since the processes under consideration are governed by the laws of physics
there is usually a lot of background knowledge, i.e. that the features cannot have unphysical values.
Incorporating this kind of knowledge into a classification scheme is very difficult, but often improves 
a classifier considerably. It can be used to exclude noisy events.
An easy way to incorporate prior knowledge is often to include the output of existing algorithms as a feature
so that only corrections have to be learned. 
In this way experience already acquired can be used in a new approach, and combined with state of the art learning algorithms.

\subsection{Problem formulation}

After the identification of the set of features $S=\{f_1,f_2,\ldots f_n\}$ 
that contains information about the class membership,
the so called feature vector (FV) $\vec{v}=(f_1,f_2,\ldots,f_n)$ can be formed by ordering the features. 
The FVs of the instances of the classification problem exist in the Feature Space $V$,
 a vector space of dimension $n$, $n$ being the number of features.
The output of a classification algorithm is 1 or -1 if the task is a two class problem. 
In that way it is a function $f$, that maps the feature space onto $\{-1,1\}$.
$f: V \rightarrow \{-1,1\}$. This function defines a border in feature space.
Depending on the side of the border the instances are, the classifier assigns a class.
The easiest kind of classification is of course linear classification where the borders are straight
lines. This is the case of a cut based analysis.
The advantage here is that the cuts are easy to interpret. 
Every algorithm that can separate two classes can readily be used for multiple classes by 
building one classifier for each class. 
That is done by training the algorithm on the training set. 
If the algorithm has several parameters which have to be tuned, 
additionally the so called validation set is used, and the performance is tested on the test set (see sect \ref{overtrainingSect}).

\section{Data Preparation}

\begin{figure}
\centerline{\includegraphics[scale=0.3]{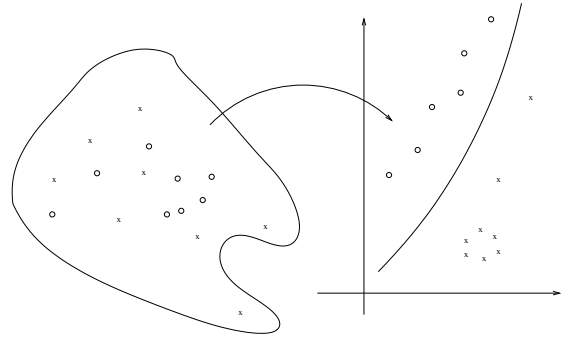}}
\caption{The problem is to separate crosses and circles. For that they have to lie in a common vector space, the feature space.
Simple Layout in feature space can eliminate the need for complex classification algorithms.}
\label{fvLayout}
\end{figure}

Fig. \ref{fvLayout} illustrates, that it is desirable to process the data such that the layout in feature space (FS) is reasonable.
i.e. the topology in FS is simple.
Usually it is easy to improve the separability of the data by preprocessing, which often renders
the use of a complex algorithm for classification superfluous.
The following sections describe several simple actions which should be taken as a first step.

\subsection{Missing Values}
An instance can only be classified or used for training of a classification
algorithm, if it is an element of the feature space. 
Therefore missing values, that would reduce the dimensionality of the feature vector have to be filled.
An intuitive choice is to use values at the borders of the range of values.
Maintaining the view of a border in feature space that is defined by the algorithm,
it is obvious that this is not necessarily appropriate.
By choosing extreme values during training the problem occurs, that an algorithm tries to place the border 
such that this instance is correctly classified. Especially if the border is a smooth function
this can have many undesirable effects, 
in particular, since the knowledge of the unphysical nature of the value is available beforehand.
During classification an extreme value can place the feature vector in a part of the feature space with 
few training examples which makes a correct classification harder.
Thus it is desirable to place FVs in densely populated part of the FS, such that the other features can determine the class, for example by using the mean value for the missing value. 
This is obviously the simplest choice. 
There exist more elaborate ways, e.g. estimating the value from the distribution of the rest of the FV by the Expectation-Maximization algorithm \cite{missValEM} or taking the mean of the nearest 
neighbors in FS where the distance is measured by the other attributes.
An introduction can be found in \cite{patClas}.


\subsection{Discretization}
\label{sectionWithDiscretization}
Many well-known data classification algorithms require discrete data. 
This includes trees but also rule finders and ultimately cut based analysis.
Given data that take on values from a continuous range, the question imposes itself, what the  
optimal way of discretization is, e.g. by trying different cuts.
The straightforward way is to choose equal steps. If the data is not distributed equally 
in the given range, it is better to use a spacing such that the number of data points in each bin
is similar.
But since these cuts should be used in the end to distinguish one class of data from another 
it is advisable to choose the cuts in such a way that it is still possible to decide on the class membership
given the discretized attribute. 
Ideally that would mean, that in each bin there are instances of only one class. 
This might, however, generate too many possible divisions.
Common modern algorithms thus employ a strategy where the the data is sorted, the possible cut-points
are determined, the data is divided at the optimal cut-point given a specific criterion and the
discretization is halted if a stopping criterion is fulfilled.
Possible cut-points are the points between label changes, the optimality criteria can 
be similar to the feature selection criteria entropy and information gain that will be talked about later in this lecture,
because the choice of discretization steps implies the choice of new features.
The stopping criterion is often mdl (minimal description length). 
For details on popular algorithms see:
\cite{FayyadAndIrani}, \cite{Kononenko}.

\subsection{Normalization}

Data normalization can be achieved simply by scaling the data to a specific range, e.g. [-1,1], 
or fixing variance to one and the mean to zero for each feature.
Another way is to use the softmax algorithms \cite{patClas} which scales data close to its mean value almost linearly, but
includes a smooth nonlinearity further away.
Data normalization simplifies the search for optimal parameters during classification. Widely different ranges for
different features can lead to numerical instabilities during computation and different weights for 
different features.

\subsection{Outlier Removal}
Outliers are a problem often encountered in physics measurements. 
They can be caused, for example,by electronics noise or mislabeling. 
Many outliers can significantly degrade the performance of classifiers, so treatment of those should be 
taken seriously. If an algorithm is used which scales the data automatically, the presence of outliers can 
concentrate the meaningful training examples in a small range of the allowed values. 
Some strategies can be found in \cite{robStat}, however, it should be kept in mind that outlier treatment also depends on the problem at hand.

\section{Feature Selection}
This section will deal with some basic aspects in feature selection.
The first topic that will be introduced is decorrelation for which the motivation is twofold.

Firstly, the assumption of statistical independence facilitates the classification task at hand. After decorrelation
it is possible to consider each feature independently.
 Simple Bayesian schemes in principle even require independent features.
Secondly, strongly correlated features can be combined, which leads to a reduction of the dimensionality of the feature space.
Having this dimensionality as low as possible is very important, since the volume of the feature space grows exponentially
with each dimension, which in turn means that the density of training examples reduces exponentially. Of course this makes
discrimination harder. 
In addition the computing time of algorithms grows strongly with the number of dimensions. This is called the curse of dimensionality.
In this part of the lecture it will be shown how after decorrelation the variance of a feature can be used as a measure for the importance.
Other measures are the information gain and significance of a feature by $\chi ^2$ which will also be introduced.

\subsection{Decorrelation}
\begin{figure}
\centerline{\includegraphics[scale=0.3]{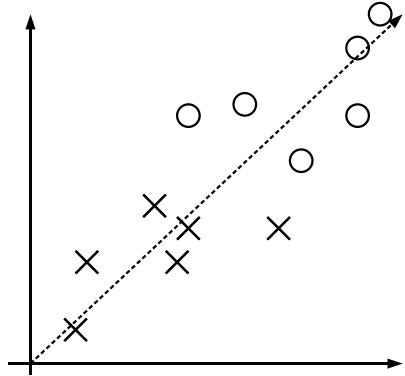}}
\caption{Principal component transformation: the data can be described in a rotated coordinate system, the dashed axis, with fewer dimensions.
One instead of two.}
\label{kltPic}
\end{figure}

The problem of reducing the dimensionality of the feature vector by making a minimal 
error is of interest for many applications. Consequently, solutions for this problem have been developed. 
The most widely known is the principal component analysis or Karhunen-Loeve-Transformation (KLT). \cite{kltCite}
The idea of the KLT is that for the removal of a dimension with the minimal error, 
it is necessary to remove the directions in which the data has the minimal variance. If only linear transformations
of the FS are allowed, the solution is unique.
An example is shown in the data distribution in Fig. \ref{kltPic}. 
There the information is mainly contained in one direction. Describing a two dimensional vector only with the component in
the direction of the maximum variance leads to the smallest expected error which can be achieved with a
linear transformation:

 A feature vector is denoted by $\vec{v}=[f_1,f_2,...,f_n]$, the  correlation matrix of the feature vectors by
 $R_{\vec{v}\vec{v}}=E\{\vec{v}\vec{v}^T\}$. Then the unitary transformation $A$ is constructed such that the feature vector after application
of the transformation becomes
$\vec{w}=A^T\vec{v}$ and the correlation matrix is given by
$R_{\vec{w}\vec{w}}=E\{\vec{w}\vec{w}^T\}=E\{A^T\vec{x}\vec{x}^T\}=A^TR_{\vec{x}\vec{x}}A=\Lambda=diag\{\lambda_1,\lambda_2,\ldots,\lambda_n\}$.
Thus the feature vectors $\vec{w}$ are now uncorrelated.

Assuming that the Eigenvalues are ordered according to their size, the KLT has some desirable features (proof given in \cite{kltCite}):
\begin{itemize}
\item $A$ consists of Eigenvectors of $R_{\vec{v}\vec{v}}=E\{\vec{v}\vec{x}^T\}$,
\item it can be shown that the expected error is $E\{\vec{v}-\vec{w}\}=\sum_{i=m+1}^n \lambda_i$,
where the $m$ first dimensions, corresponding to the variances and eigenvalues of $R_{vv}$ $\lambda_1\ldots \lambda_m$
are omitted,
\item the projection onto a subspace spanned by the first $m$ Eigenvectors is optimal,
\item the sum of the eigenvalues, which are the variances in the direction of the corresponding eigenvector, 
can be used as a measure of the information contained in this feature,
\end{itemize}
and some undesirable:
\begin{itemize}
\item it is data dependent, since the covariance matrix has to be estimated from data.
Also, there has to be enough data to be able to invert the matrix,
\item there exists no fast algorithm,
\item it is not optimized for class separability, 
this is incorporated into Fishers linear discriminant \cite{fisherCite}, which will not be discussed here.
\end{itemize}

Since the KLT is not optimized for class separability it is ad hoc problematic to identify the 
direction of maximal variance with the direction of maximal importance. 
Possibly the direction for optimal separability is one of small variance. 
However, even if the full dimensionality is kept, the decorrelation is already of importance for many tasks.
For example maximum likelihood methods assume that the data is not correlated.
But the KLT is only one transformation that can make the feature space more amiable for classification.
Depending on the application, other representations may be useful. Popular examples are the Fourier transform 
or the Haar transform which leads to robustness against transformations of the data by integration over these.

\subsection{Significance of an attribute for classification}
The best feature for classification is obviously one that already determines the class membership of the example.
Then the information gained by knowing the correct classification equals zero, since 
the outcome is already known, whereas without knowing the value of the feature the information gain was big, since the outcome was completely uncertain.
This is an extreme example, and in most of the cases finding a feature like this is impossible but
the generalization demonstrates that the better the feature is, the more the outcome of the classification 
is fixed. 
The actual classification will not be surprising, if the probability of a specific outcome is already 99\%
after the knowledge of a certain attribute. 
This means that the quality of the attribute can be measured in a way by the uncertainty of the outcome after knowing the attribute compared to the one before knowing it.
To put this in a less colloquial language, the significance of an attribute 
for classification can be measured by computing the information gained from the knowledge of the value of the attribute, the so called information gain.

In information theory the information contained in the outcome of a random draw is defined as
\begin{equation}
\mathrm{Inf}(P(v_1),\ldots,P(v_n))=\sum^n_{i=1}-P(v_i)log_2P(v_i)
\end{equation}
if there are $n$ possible outcomes $v_1,\ldots,v_n$, each with a probability $P(v_n)$.
The unit of information is a bit. 
During classification a feature vector is randomly drawn and the class membership is to be determined.
Assuming there are only two different classes with equal probability ($P(c_1)=\frac{1}{2},P(c_2)=\frac{1}{2}$) the information contained in the outcome
is thus 
\begin{equation}
I(\frac{1}{2},\frac{1}{2})=-\frac{1}{2}log_2\frac{1}{2}-\frac{1}{2}log_2\frac{1}{2}=1 \mathrm{bit}
\end{equation}
This could be for example the toss of a fair coin. The information contained in the outcome is 1 bit, as expected.
If the coin is not fair, and it is known for example that the outcome will be in 99\% of the cases heads, the information
contained in the outcome will be less
\begin{equation}
I(\frac{1}{100},\frac{99}{100})=0.08 \mathrm{bit}
\end{equation}
and the information gain in bit of this knowledge will be $1-0.08=0.92$.

Fig. \ref{diffFSDivides} illustrates how the knowledge of one discrete 
attribute devides the Feature Space in different partitions in a classification problem. 
Each of the partitions has a different probability for the two classes.
Before, without knowing anything, the information contained in the outcome of the classification of one of the examples was 1 bit,
assuming that there are the same number of examples for both classes.  
With the prior knowledge of a feature, the partition of the feature space in which the example that is to be classified lies is known.
Depending on the distribution of classes in this partition the classification provides significantly less information. If the partition contains 
only one class, the information gained from knowing the class label in addition to the partition is even zero.

To compute the information left after knowing in which partition the example lies (by knowing the attribute A),
 the information
still needed to determine the classification in each partition weighted by relative number of examples in this partition can be computed:
\begin{equation}
  \mathrm{Remainder}(A)=-\frac{\sum_{i=1}^v p_i+n_i}{p+n}I\left(\frac{p_i}{p+n},\frac{n_i}{p+n}\right)
\end{equation}
For a set containing $p$ positive and $n$ negative examples the computation of the information is straightforward:
\begin{equation}
I_{\mathrm{Init}}\left(\frac{p}{p+n},\frac{n}{p+n}\right)=\frac{-p}{p+n}\log_2\frac{p}{p+n}-\frac{-n}{p+n}\log_2\frac{n}{p+n}
\end{equation}
Thus the information gain of $A$ is $I_{Init}-\mathrm{Remainder}(A)$.
Information gain is a popular measure for the significance of an attribute for classification, the higher the value the
better the attribute.
Given a set of possible attributes,  meaningless attributes can be excluded by computing the information gain for each.
This value can also be used to order the attributes in the nodes of decision trees.

\subsection{Determining Significance by $\chi^2$}
\begin{figure}
\centerline{\includegraphics[scale=0.3]{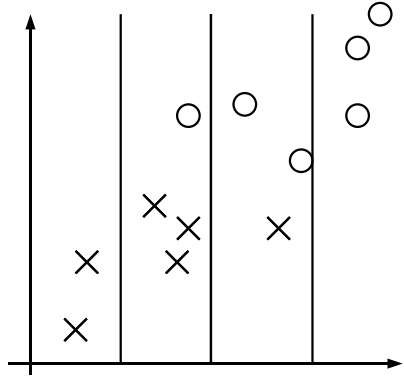}}
\caption{A discrete attribute divides the feature space into different partitions. For a significant feature the distribution in 
each  will deviate significantly from the expected value given the distribution of feature vectors without the knowledge 
of the feature.}
\label{diffFSDivides}
\end{figure}

Another ansatz for the determination of the significance of an attribute is the question if the probability distribution
function (PDF) changes significantly in one of the partitions of Fig. \ref{diffFSDivides}.
For  the difference between the expected number of positive and negative examples given the PDF can be computed
 with the actual number of positive and negative examples.
This is done by computing the $\chi^2$ between the null hypothesis, the feature is irrelevant and the PDF will be unchanged,
and the hypothesis that the feature is relevant and the PDF is changed.
If the impact of the attribute on the PDF is big, the $\chi^2$ will also be big. Thus the significance of the attribute
can be deduced from  the $\chi^2$ that can be computed in a straightforward way if the attribute splits the set of examples 
l-fold. First the expectation values for the number of positive and negative examples in set $i$ ($\hat{p_i}, \hat{n_i}$) is computed:
\begin{equation}
  \hat{p_i}=\frac{p\cdot(p_i+n_i)}{p+n}, \quad \hat{n_i}=\frac{n\cdot (p_i+n_i)}{p+n}
\end{equation}
if the attribute is irrelevant the deviation $D$ should follow a $\chi^2$ distribution:
\begin{equation}
D=\frac{\sum_{i=1}^v(p_i-\hat{p_i})}{\hat{p_i}}+\frac{\sum_{i=1}^v(n_i-\hat{n_i})}{\hat{n_i}}
\end{equation}
The significance of the attribute then follows from the actual $\chi^2$.

\section{Basics in Statistical Learning, Cost functions}
\begin{figure}
\centerline{\includegraphics[scale=0.3]{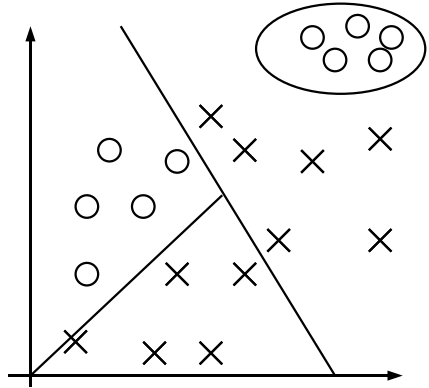}}
\caption{Possible forms of a classifier: linear and mahalanobis distance}
\label{diffClassifications}
\end{figure}

This section introduces some basic classification schemes which follow from basic statistics.
The problem to be solved is illustrated in Fig. \ref{diffClassifications}. Given the feature space 
with the training examples the task is to find the partition of the space that minimizes the classification error.

Usually the more general concept of classification cost is used. The definition of an appropriate cost function then allows 
to weight different errors by their importance.
In basic classification  usually the cost $C$ is defined as the fraction of misclassified examples:
$C=\frac{1}{n}\sum_{i=1}^n \|f(v_i)-\frac{1}{2}\mathrm{classlabel}(v_i)\| $
where $f$ is the classification function giving the assigned classlabel $1$ or $-1$ for the test examples $v_i$.
This is not always appropriate for physics problems, since here the goal is often to maximize a figure of merit.
If $S$ is taken as the signal and $B$ the background this has for example the form $S\cdot\frac{S}{S+B}$.
More problems with the use of basic cost function occur if the number of training examples in each class is not weighted 
correctly.  
Therefore statistical learning toolboxes usually give the user the choice of the cost function for example in the form 
of a matrix assigning a different cost for each possible confusion.

\subsection{Dealing with physics data}
Often, when classifying physics data, it is a  problem to find appropriate 
training examples. Either the examples are from simulated (MC) data, or from real data, but then the labeling is most
of the time not 100\% perfect. 
In both cases, it is advisable to study the effects of this on the performance of the classifier. If it performs good 
on the training data but gets worse if there are some training examples with the wrong label or are a little bit shifted in 
feature space, one cannot expect a good performance on real data. 
Most classifiers also have parameters that control the generalizing power. This is done to avoid overtraining (see section \ref{overtrainingSect}) . 
For example rule based classifiers or decision trees have a pruning stage. The algorithm should be used in a way
that maximizes its ability for generalization.

\subsection{Classification with Bayes theorem}
The most basic, and optimal, classifier tries to maximize the probability for correct classification. It thus 
chooses the class $\omega_k$ according to
\begin{equation}
\max_k{P( \omega_k | \vec{v}  )}
\end{equation}
This is the so called maximum a posteriori (MAP) classifier. Given evidence, it gives the probability 
of each possible classification.
However, to evaluate this expression  the conditional probability distribution 
 $P(\omega_k |\vec{v})$ has to be known for each possible feature vector $\vec{v}$ which is usually impossible.
Therefore the formulation is recast with the help of Bayes theorem. This says it is possible to express 
the a posteriori probability in terms
of the probability of the values for the feature vectors given the class
  $P(\vec{v}|\omega_k)$ and the a priori probabilities for each class $P(\omega_k)$ and the probability for
the specific feature vector $P(\vec{v})$ as:
\begin{equation}
P\left(\omega_k |\vec{v}\right)=\frac{P\left(\vec{v}|\omega_k\right)P\left(\omega_k\right)}{P\left(\vec{v}\right)}
\end{equation}
The standard example for this theorem, given in the following, comes from medicine. 
Given certain symptoms, the MAP classifier gives the probability for an illness. 
After the reformulation with the help of Bayes theorem it can be seen that this depends heavily on the a priori probability 
for having the illness. Given a rare illness that only occurs once in ten thousand cases, its probability can be computed
 after a positive test.
Given that the test is correct 98 \% of the time, so for a rare illness 
approximately for 2 \% percent of all cases it gives a positive result, the probability for the illness even after 
positive test is, $98\% \cdot 2\% \frac{1}{10000}=0.5\%$. 

By using the Bayes theorem surprisingly good results can be achieved on many datasets.
The classification question is reduced to the question
\begin{equation}
  \frac{P\left(\vec{v}|\omega_j\right)P\left(\omega_j \right)}{{P(\vec{v})}}
 \stackrel{?}{\geq} \frac{P\left(\vec{v}|\omega_k\right)P\left(\omega_k\right)}{P(\vec{v})}
\end{equation}

\begin{figure}
\centerline{\includegraphics[scale=0.5]{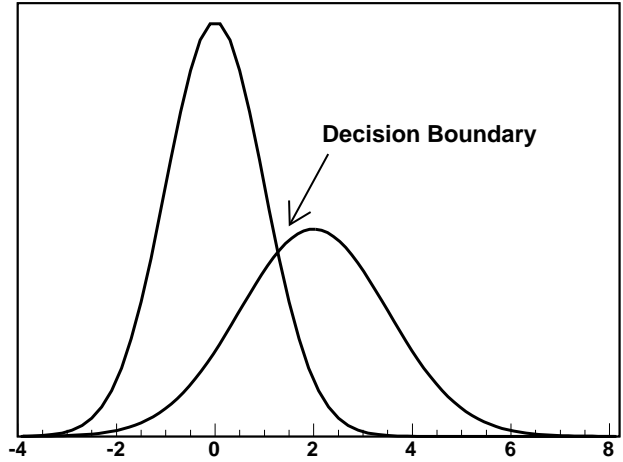}}
\caption{Probability vs. feature value for two classes for a 1-dimensional problem, assuming normal distributions.}
\label{twoGaussians}
\end{figure}

For actual classification it is necessary to make certain assumptions regarding the occurring probability distributions.
The natural choice for the conditional PDFs is that the features given a class are normally distributed around 
the mean value for that feature given the class.  

\begin{figure}
\includegraphics[scale=0.3]{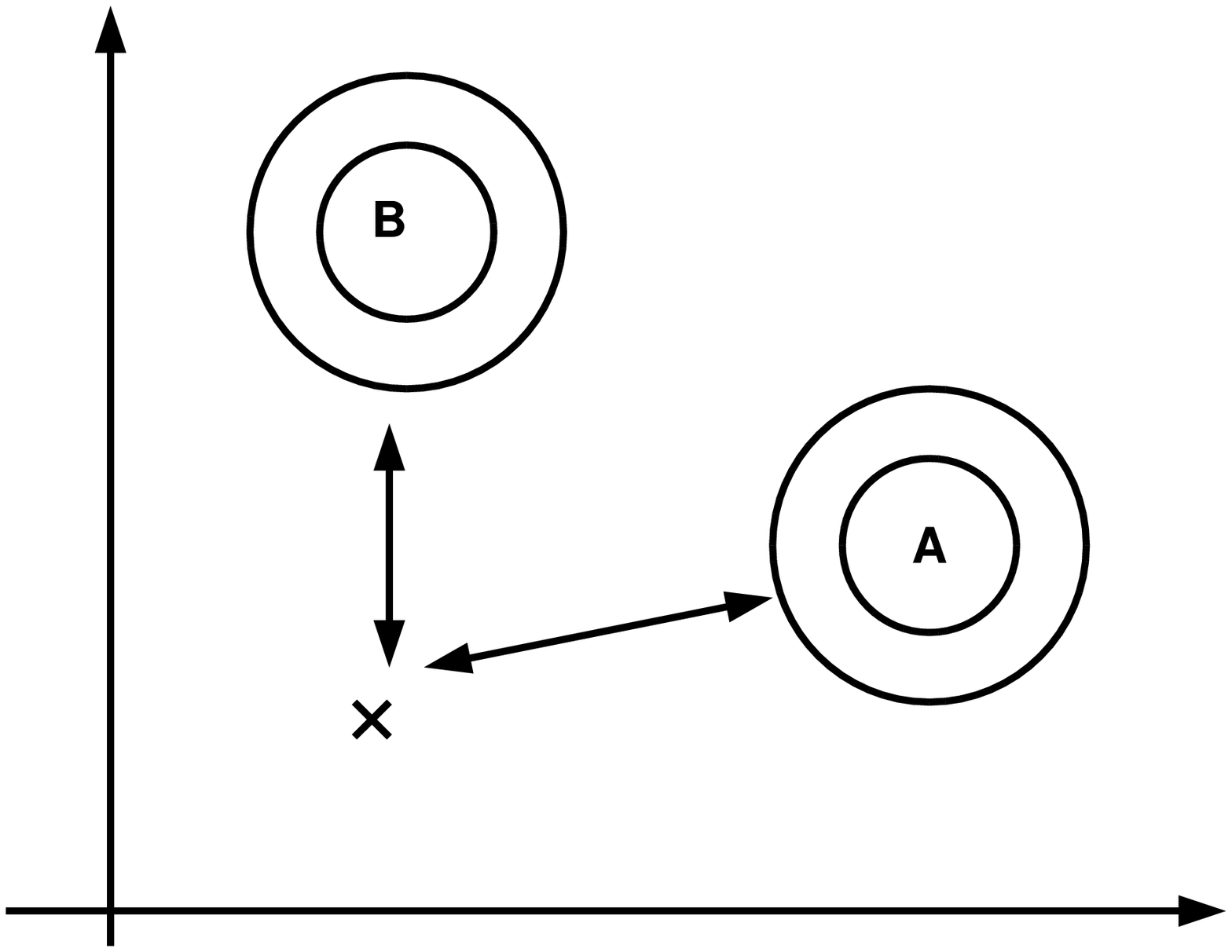}
\includegraphics[scale=0.3]{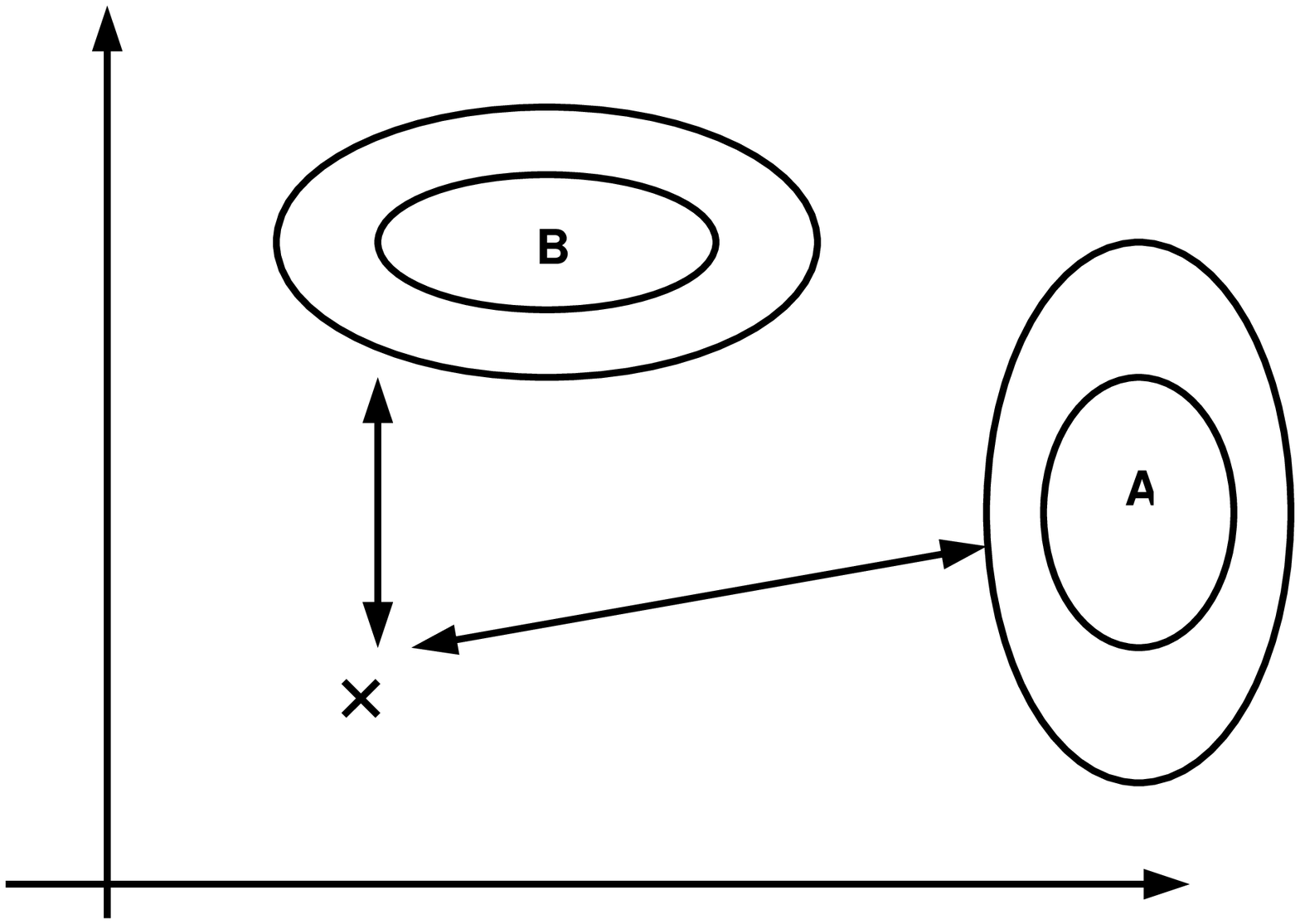}
\caption{Euclidean and Mahalanobis distance of one point to the classes A and B. The lines of equal distance are circles and ellipses, respectively.}
\label{figWithEuklidMahalDist}
\end{figure}

Fig. \ref{twoGaussians} shows the one-dimensional case. 
Here the classification is done by simply evaluating each PDF at the value of the feature and choosing the class that has the higher
probability there.
The a priori probability of the classes can in principle be computed by sampling.
But it is also possible to use the assumption of equal probabilities $P(\omega_k)$ for all classes
and uncorrelated features, which can be achieved by the application of the KLT, each distributed with the same variance $\sigma$, if this is reasonable for a given problem.
The resulting classifier is called Maximum Likelihood Estimator (MLE).
Under the assumption of uncorrelated entries in the feature vector, the probability 
$P(\vec{v})$ is just given by the product of the probabilities of the entries, which have a Gaussian distribution around
the mean value of the given class $\omega_k$, $\mu_k$
Thus:
\begin{eqnarray}
P(\omega_k)\propto e^{-\frac{(v_1-\mu_k^1)^2}{\sigma^2}}\cdot \ldots \cdots e^{-\frac{(v_n-\mu_k^n)^2}{\sigma^2}}&=&
e^{-\frac{\|\vec{v}-\mu_k\|^2}{\sigma^2}} \\
\Rightarrow P(\omega_k|\vec{v})\stackrel{?}{\geq}P(\omega_j|\vec{v}) &\Leftrightarrow& \|\vec{v}-\mu_k\|\stackrel{?}{\geq}\|\vec{v}-\mu_j\|
\end{eqnarray}
So the classification is reduced to the computation of the euclidean distance of the feature vector to the mean of the class.
In a two dimensional feature space the lines of equal probability are circles. If the covariance matrix is known and if it is 
not equal unity it is possible to use a different metric, the mahalanobis distance. This incorporates the covariance matrix $K$ by
using 
\begin{eqnarray}
P(\omega_k)&\propto& e^{-(\vec{v}-\mu_k)K^{-1}(\vec{v}-\mu_k)^T}\\
K&=&E\left\{(\vec{v}-\bar{v})(\vec{v}-\bar{v})^T\right\}
\end{eqnarray}
where $E$ is the expectation value and $\bar{v}$ is the mean of $\vec{v}$.
The distance induced by this is the so called Mahalanobis distance $D_M$, the distance between
two points $\vec{v}$, $\vec{w}$ in a vector space can thus be computed by
\begin{equation}
D_m^2(\vec{v},\vec{w})=((\vec{v}-\vec{w})K^{-1}(\vec{v}-\vec{w})^T
\end{equation}
and with this notation
\begin{equation}
P(\omega_k|\vec{v}) \stackrel{?}{\geq}P(\omega_j|\vec{v}) \Leftrightarrow D_M(\vec{v},\mu_k) \stackrel{?}{\geq}D_M(\vec{v},\mu_j)
\end{equation}
As illustrated in figure \ref{figWithEuklidMahalDist}, the lines of equal probability in a two dimensional space are now elliptic.

\subsection{Rule based classification for automated cut finding}
In this section automated cut finding with the RIPPER algorithm is introduced, which was not included in the original lecture.
Many statistical learning tools also incorporate rule based classification, which can easily translated
into a cut based approach. The advantages are of course that simple rules can be interpreted more easily,
interpretation of the data is facilitated and most people have more confidence in cuts than in black box algorithms 
like neural network.
The RIPPER algorithm was introduced by William W. Cohen \cite{RipperPaper}. 
and has been known for a long time in the machine learning community.
A similar approach is realized by decision trees which are not discussed here.
In general, rule based algorithm need discretized attributes. Thus they have to be combined with one of the strategies for
discretization. Examples were covered in section \ref{sectionWithDiscretization}.
The RIPPER algorithm  builds a ruleset $RS$ for classification. The ruleset is initialized with the empty set in the beginning, 
$RS=\{\}$. The goal of the algorithm is to grow rules by adding conditions (connected by the logical and) so each rule covers
a convex space in Feature Space. An example is positively classified if any rule matches. 
Thereafter comes the building stage. This consists of repeatedly executing the following steps until a stopping criterion is reached:

\begin{itemize}
\item grow phase - grow one rule by greedily adding new conditions according to the highest information gain of the attribute
until the rule is perfect. I.e. it covers all positive examples and no negative.
\item prune phase -  tune the rules to the training set, since they are 100\% perfect. This is the problem of 
overtraining (see section \ref{overtrainingSect}). It is thus necessary to prune the set. 
All finite sequences of conditions are considered, if a sequence covers $p$ positive and $n$ negative examples, 
and the value $2p/(p+n)$ is below a certain threshold, it is pruned. This makes sure that the rules are not customized to 
single examples. The threshold can be chosen according to variance in the data.
\end{itemize}
The algorithm stops if the rule lengths are too different or all positive examples are covered or the error rate is too high.
In the end, the initial ruleset is optimized. 
The set of rules obtained in this way can then be easily translated in cuts.
Experience shows that taking care during the discretization step is very important, because measurements are not spread evenly over
the allowed range of values and if the discretization is too coarse in regions where they are dense and small differences matter,
the best algorithm cannot achieve a good performance.

\subsection{Performance estimation}
\label{overtrainingSect}
\begin{figure}
\centerline{\includegraphics[scale=0.6,trim=50 0 50 250, clip]{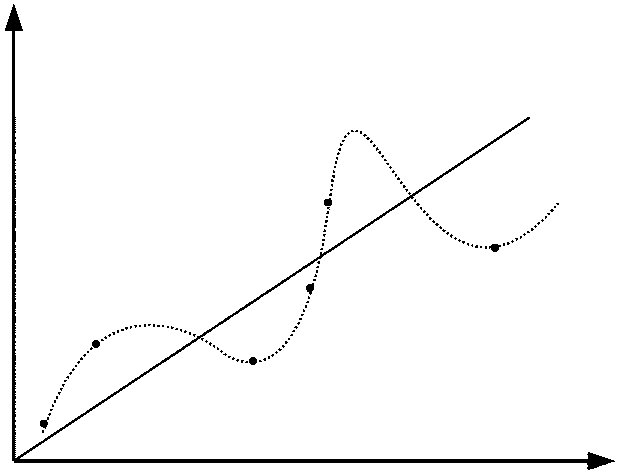}}
\caption{Possibilities to fit a function with different complexities.}
\label{figWithOverfit}
\end{figure}

The performance of an algorithm on a training set is not necessarily a good indicator of its performance on real data.
A perfect performance often comes from a specialization on the given training data. This leads to a loss of generality 
and thus most of the time to a poor performance on real data.
This problem is known as \begin{bf} overtraining\end{bf} and is always an issue in machine learning and consequently 
all algorithms have measures 
to control the generalization capability.
Support Vector Machines (SVMs \cite{svmCite}) have a solid theoretic background by which, in principle, the generalization ability can be computed.
Other algorithms have more ad-hoc parameters, trees and rule finders use pruning, neural networks often use weight decay.
The principle is that of Occam's razor. This states that it is always advisable to use the simplest theory that describes the data. 
Especially with little training data or with data that is uncertain, as it is often the case with physics data, 
a simple model should be used. This follows also from a kind of Bayesian reasoning: a complex theory, a theory that takes a lot of bits to describe,
is more unlikely, so more evidence is needed to support it. See for example Fig. \ref{figWithOverfit}. 
Here the training performance of 
a linear function is worse, but looking at the data it seems to be the correct description, especially if expected scattering of the data is considered.
There exists an elaborate theory that connects the predictive power of a function and the training error to the expected error, the Vapnik-Chervonenkis theory,
and SVMs \cite{svmChapter} are build according to this principle, but to explain this would be too much detail.
It is enough to keep this problem in mind, especially in physics, and evaluate the performance in a way that includes the generalization ability.
For this an independent test set is needed, that had nothing to do with the training of the algorithm.
If the algorithm has also meta parameters, which have to be determined in an optimal way, in principle three sets are necessary.
One for training the algorithm with a specific set of parameters, one to determine the optimal parameters, the so called validation set, and one to evaluate the whole algorithm
with the chosen parameters. 
Otherwise there is also the risk that parameters are tuned to the training set.  
It is also advisable to incorporate a bit of scattering in the training examples to see how the algorithm performs.
To compute the expected error, it is common practice to either just split the data 3 ways or do the so called \begin{bf}n-fold cross-validation\end{bf}.
Here the data is split in $n$ folds, then the training is performed on $n-1$ sets and the remaining set is used for testing.
This is done {\bf n} times and the expected error is computed as the mean error. 
It can be shown  that the higher $n$, the better the estimation is.
The most extreme case being the leave-one-out (loo) error, where the sets only have one element \cite{errorEstimates}.

\section{Tools}
Getting to know existing tools is very important for two reasons. 
Firstly, of course, it is not necessary to invent anything that has already been implemented 
and there are a lot of existing toolboxes which can be used.
The second reason is, that toolboxes provide the opportunity to try many different strategies which enables
the user to get a feeling what might work for a given problem.
This is one of the most important things that one has to do while implementing a successful 
strategy for solving a classification task. 
Therefore two toolboxes are presented in this section that make it possible, even for a rather unexperienced user,
to try out different algorithms for data normalization, preparation and classification, the Weka toolbox \cite{weka} and the Toolkit for Multivariate Analysis (TMVA: \cite{tmva}).
The Weka tool has the advantage of implementing almost every popular algorithm and combining this in a unified graphical user 
interface, providing a workplace where different algorithms can be combined and tested.
In particular, all algorithms presented so far are implemented, so that it is possible to try them right away.
On the other hand  TMVA developed using the ROOT software, is particularly aimed at analyzing particle physics data. 
For example it allows the definition of the signal significance $S^2/(S+B)$ as the figure of merit to be optimized and is fully integrated in the ROOT
environment. 
As of this writting, TMVA is still under heavy development but stable and usable. It is part of ROOT, but since its development cycles are currently
faster, it should be downloaded separately.
Easy to use graphical user interfaces allow to choose optimal working points, inspect the data and compare the performance of different classifiers.
It contains a scripting as well as a C++ interface. There is also a complete users guide available.

\end{document}